# Predicting Emotional Volatility Using 41,000 Participants in the United Kingdom

**Authors:** George MacKerron[1], Nattavudh Powdthavee[*2]

**Affiliations**:

[1]University of Sussex, Falmer, Brighton, BN1 9RH, UK

[2]Nanyang Technological University, School of Social Science, 48 Nanyang Avenue, 639818, Singapore

*Corresponding author. Email: n.powdthavee@gmail.com

**Keywords**: emotional volatility; asymmetry; intrapersonal variability; happiness; experience sampling method

**Abstract**

Emotional volatility is a human universal. Yet there has been no large-scale scientific study of predictors of that phenomenon. Building from previous works, which had been ad hoc and based on tiny samples, this paper reports the first systematic estimation of volatility in human emotional experiences. Our study draws from a sample of intrapersonal variation in moment-to-moment happiness from over three million observations by 41,023 UK individuals. Holding other things constant, we show that emotional volatility is highest among women with children, the separated, the poor, and the young. Women without children report substantially greater emotional volatility than men with and without children. For any given rate of volatility, women with children also experience more frequent extreme emotional lows than any other socio-demographic group. Our results, which are robust to different specification tests, enable researchers and policymakers to quantify and prioritise different determinants of intrapersonal variability in human emotions.

All humans experience daily variability in emotions. Yet, despite well-being metrics becoming much more readily available in many large-scale surveys, there appears to have been no systematic, large-scale scientific study of the predictors of that phenomenon. Much of the focus of previous studies has been on analysing the mean scores of evaluative well-being, which is a person's summary evaluation of their life, as the statistic of interest. These studies, often with more than 10,000 individuals in the sample, have produced many interesting findings, including the evidence that unemployed people report substantially lower overall life satisfaction, on average, compared to the employed[1, 2], and that economic growth does not necessarily lead to an improvement in the average happiness of citizens in the long run[3, 4]. The recent availability of experienced well-being data in large-scale surveys has also given researchers enough degrees of freedom to systematically study what matters to how a person feels at the moments of their life and the extent to which these predictors are different from those that determine how satisfied they are with their life overall. We now know, for example, that income is a much better predictor of evaluative well-being than experienced well-being[5, 6], while a rise in income inequality has a similar depressive impact on both dimensions of well-being[7].

One of the most important implications of large-scale studies of mean evaluative and experienced well-being data is that they enable researchers to quantify and prioritise what matters the most on average to citizens' well-being, which, in turn, inform policymakers on which policies to pursue that would be 'bang for the buck' at improving societal welfare[8]. For instance, given the evidence that an increase in a country's unemployment rate depresses people's evaluative well-being almost twice as much as a rise in inflation[9], the country's central bankers may find it more optimal to trade-off a rising in prices for a lower overall unemployment rate. In a study that uses over 8,000 observations of British adults, researchers find maternal mental health in a child's early years to be one of the most important predictors of adult life satisfaction at age 34[10]. This finding informs policymakers that an intervention on new mothers' mental health may be more cost-effective at raising individual well-being in adulthood than other interventions. Estimates from running multiple regression analysis on evaluative and experienced well-being data have also been used to estimate 'shadow prices' – i.e., the additional income required to fully compensate a drop in well-being from a change in life event *X* – for things that are not priced in the market, such as air pollution[11], ill-health[12], and even hedonic damages from bereavement[13], which can then be used to guide policy decisions.

However, despite a wealth of literature in psychology around the measurement, dynamism, and emodiversity of human emotions and how emotional variability contributes to a person's psychological well-being[14-18], large-scale studies of predictors of volatility in experienced well-being are virtually non-existent. Previous investigations of the potential determinants of emotional variability such as gender, income, marital status, and age tend to be *ad hoc*, focused on only one predictor at a time, and often based on extremely small samples[19-26]; see Table 1 for a summary of some of the notable studies. Different studies also used different samples and data collection methods, thus making systematic comparisons between predictors particularly challenging. As a result, we currently have little scientific insights into the relative importance and trade-offs between different predictors of emotional variability between- and within-person, which, in turn, means that there is little opportunity to incorporate findings around the determinants of emotional variability into the well-being and public policy discussions.

Data availability explains the lack of empirical evidence in this area. Large-scale, longitudinal surveys such as the Understanding Society (UKHLS) and the German Socio-Economic Panel (SOEP) interview the same individuals only once a year. The low-frequency nature of these data only allows researchers to study evaluative well-being volatility over long periods of time, e.g., over ten years or more[27], but not over much shorter periods, e.g., weeks or months. Although more appropriate, the day reconstruction method (DRM), which asks individuals to reconstruct their activities and experiences of the preceding day[28], typically consists of only one day of well-being volatility data per person. The DRM is also subject to both memory and present biases[29]. The experience sampling method (ESM), which requires participants to provide self-reports of their activities and emotions multiple times per day, is the most suitable data collection method for this type of research. However, such high-frequency data are rare and often have a very small sample size. Moreover, because of the demanding nature of the data collection process, they tend to target specific groups of individuals, such as university students[30], thus making them unsuitable for studying how emotional variability varies across people with differing socio-demographic statuses.

As an empirical matter, the ESM data must have several special features to make a significant advancement in this research space:

(i) the sample must be sufficiently large to allow enough degrees of freedom for multiple parameters to be estimated in a regression model;
(ii) individuals in the sample must be followed over a reasonably long period, so that plenty of intrapersonal variation in experienced well-being is available;
(iii) the sample should be reasonably representative of the adult population;
(iv) a set of important variables, such as income and marital status, must be available in the data set, so that confounding influences can be differenced out and comparisons across different estimates can be made.

No study of this type has apparently been published with emotional variability as the outcome variable of interest.

To fill this research void, this study uses Mappiness data[31], one of the largest ESM datasets in the world, with over three million responses from more than 41,000 UK participants. After downloading and signing up to the Mappiness app, participants, who are iPhone users, are initially asked about their life satisfaction: "How satisfied are you with your life as a whole nowadays?", on a scale of 1 – 10. They also fill in answers to a set of questionnaires about their socio-demographic characteristics, including age, gender, marital status, employment status, health status, household income, and the number of children in the household. The app then pings individuals at random moments one or more times a day before asking how 'happy', 'relaxed', and 'awake' they feel using continuous sliding scales that we scale from 0 'not at all' to 100 'completely'. While users answer, location is determined using satellite positioning (GPS). All data is sent back wirelessly, anonymously, and securely to a central data store. Since its creation, researchers have used Mappiness data to analyse a variety of questions, including relationships between average moment-to-moment happiness and scenic locations[32], noise pollution[33], and work[34].

The large size, high frequency, and longitudinal nature of the Mappiness data set, gathered between 2010 and 2016, makes it ideally suited to analysing how different socio-economic variables predict intrapersonal emotional variability over short periods of time. By focusing on a simplified measure of intra-personal emotional variability or 'emotional volatility' index, $V$, which is the weekly standard deviation of moment-to-moment happiness per person[35, 36], we obtained 338,447 observations from 41,023 unique individuals over time.

This paper is the first to estimate three sets of regressions with the following dependent variables on the same data set:

- Emotional average index, $M$,

- Emotional volatility index, $V$, and
- Emotional 'skewness' or asymmetry index, which is weekly mean happiness minus weekly median happiness divided by the weekly standard deviation, i.e., $\frac{Mean_w - Median_w}{V_w}$, $A$.

For raw data distributions of the three indices, see Figs. S1A-C in Supplementary Material (SM). To make the coefficients comparable across the three regressions, we standardised all dependent variables across the sample to have a mean of 0 and a standard deviation of 1. This process allows coefficients to be interpreted as a standard deviation change in the dependent variable. We included the following independent variables in all regressions: age, age squared, gender, log of household income, variables for employment status, marital status, self-reported health, parental status, number of adults in the household, year at first response, home region, the total number of prior responses and its square, and the number of responses in the corresponding week. Given that people who reported extraordinarily high or low well-being will naturally have a lower within-person standard deviation, we also included the weekly happiness mean and its squared term as additional control variables in the volatility and asymmetry regressions. Moreover, since previous research has found that women experience significantly more time stress from parenthood than men[37], all regressions allowed for interaction effects between having children and gender. The study used a linear Ordinary Least Squares estimator with standard errors adjusted for clustering at the individual level.

**Results**

The key contribution here is that we can systematically examine the socio-economic determinants of $V$ and compare them to the standard predictors of $M$. We report the full regression results in Table 2.

**Predicting emotional means**

Regressing *M* on a set of socio-economic factors produces qualitatively similar results as those typically found in the well-being literature[38]. For example, *M* is increasing with income; a 1% increase in household income is associated with a .046 SD [95% CI: .018, .074; *p*=.001] increase in *M*. There is little statistical difference in *M* between parents and nonparents, regardless of their gender. We find some evidence that the unemployed reported lower average well-being compared to those in employment or self-employment; they reported .098 SD [95% CI: -.180, -.015; *p*=.020] lower *M* than the full-time employed. Consistent with previous findings[39], well-being average is U-shaped in age, with a minimum at 38 years old. There is evidence of a sharp decline in *M* with poorer self-reported health. We also found the intra-week emotional average to be significantly lower for those not working due to long-term sickness and higher for those who are married.

**Predicting emotional volatility**

Shifting our attention to *V*, which is our main statistic of interest, we find that income is associated with a significant reduction in well-being volatility; a 1% increase in household income is associated with a .021 SD [95% C.I.: 0.006, .036; *p*=.007] decrease in *V*. Despite finding little difference in terms of happiness between people with and without children, *V* is substantially higher among those with children in the household. Compared to men with no children, *V* is .094 SD [95% C.I.: .069, .118; *p*<.000] higher for women with no children; .040 SD [95% C.I.: .003, .077; *p*=.034] higher for men with children; and .135 SD [95% C.I.: .099, .170; *p*<.000] higher for women with children. F-tests of equality of variances also reveal that women with children experienced substantially higher well-being volatility compared to both women without children [*p* < .019] and men with children [*p* < .001].

We find some evidence that the unemployed experienced significantly lower levels of *V* compared to the employed. Students in full-time education, though reporting similar *M* as those in employment or self-employment, also experienced substantially lower *V* over short periods of time. Despite reporting similar levels of *M* as those who have never been married, those who are separated or divorced from their partners experienced around .09 SD higher *V*, on average. Marriage is positively associated with both *M* and *V*. By contrast, both *M* and *V* are significantly lower for people with long-term illnesses. There is evidence that *V* declines at a decreasing rate with age. To get some idea of the comparative differences across these socio-demographic variables, we present *z*-scores obtained from the *V* regression in Fig. 1. We also compare the *z*-scores obtained from both *M* and *V* regressions in Fig. S2 in *SM*.

**Predicting emotional asymmetry**

We next turn to the question of emotional asymmetry, *A*. For any given *V*, are there patterns in how that variability is distributed? Specifically, can we predict whether a person's more extreme reports are likely to be on the high or the low side? One can imagine that two people with the same well-being volatility can have two very different distributions; for an illustration, see Figs. S3A-B in *SM*. To address this question, we estimated *A*, in which the dependent variable is the weekly mean happiness minus the weekly median happiness divided by the weekly standard deviation, i.e., $\frac{Mean_w - Median_w}{V_w}$. This value is then standardised to have a mean of 0 and a standard deviation of 1. Here, a negative coefficient in the skewness regression implies a negative skewed well-being distribution, indicating a more frequent reporting of extremely low happiness for any given *V*. The opposite applies to a positive coefficient in the skewness regression.

The asymmetry regression, which we report Table 3, produces coefficients that are mostly statistically insignificantly different from zero, suggesting that the distribution of *V* is similar across most socio-economic groups. There are, however, some notable exceptions. For example, the coefficient for women with children is -.049 SD [95% C.I.: -.069, -.029; *p*<.000], which implies that women with children experience not only the highest *V* but also one of the most negatively skewed emotional distributions across all socio-demographic groups in our data. By contrast, individuals in full-time education have one of the most positively skewed well-being distributions, suggesting that students experienced substantially more frequent extreme emotional highs than others.

**Robustness checks**

What might explain the differences in emotional volatility and asymmetry across socio-demographic statuses? One theory is that people from certain socio-economic backgrounds have accumulated more 'hedonic capital' over the years that, consequently, enables them to become more psychologically resilient and emotionally stable over time[40]. An alternative answer might lie in the variability of activities undertaken by people from different socio-demographic backgrounds. It might be that greater emotional volatility is driven by greater variety in our daily activities. However, including within-person variability in daily activities in the happiness variability regression does little to alter the signs and statistical robustness of the socio-demographic coefficients; see Table S1 in *SM*. Differences in emotional volatility across different groups of people remain unexplained even after conditioning for the differences in the within-person variability in daily activities experienced by individuals each week.

This study is not without limitations. One objection is that, like other experienced sampling studies, the Mappiness data set is not nationally representative, and that we cannot therefore be confident of generalising our findings to the UK population. Following a guideline given by Matthew Killingsworth[41], one way to check whether our results might be generalisable is to assess whether our sample "behave like" a representative sample in terms of key variables that are shared across the current study and previous studies of representative samples. Given that we have data on life satisfaction, which is the same measure of individuals' evaluative well-being found in the nationally representative Understanding Society (UKHLS), we can assess whether its key predictors are qualitatively similar across the two samples. Regressing life satisfaction on socio-demographic status does produce qualitatively similar findings as those obtained in studies that used nationally representative data sets in their analysis[38]; for the regression results, see Table S2 in *SM*. For example, life satisfaction is positively and statistically significantly correlated with incomes and is U-shaped in age. The unemployed report substantially lower levels of life satisfaction compared to the employed, while women are marginally more satisfied with life than men. These results offer us some reassurance that we can reasonably expect the findings for emotional volatility to generalise as well.

Second, the results are arguably confounded by unobserved time-invariant characteristics such as personality traits that correlate with both the socio-demographic variables and happiness variability. For example, one could imagine that more conscientious individuals might earn higher incomes and, at the same time, have the tendency to report a relatively stable well-being over short periods of time. However, since socio-economic characteristics in an experience sampling study very rarely change while the participants are enrolled in the study, we cannot use a fixed effects estimator to prevent the time-invariant factors from confounding our results. To test the extent of this problem, we employed a fixed effects filtered model[42] that allows us to filter out the fixed effects while preserving the variables that do not change (or change rarely/slowly) over time, like gender and education. We find that the overall results – see Table S3 in *SM* – continue to hold even after conditioning for individual fixed effects, alleviating the concern that our findings are biased by unobserved individual differences.

Third, the relationships estimated in this study are not causal. Take the negative relationship between income and well-being volatility, for example: a highly plausible explanation is that poorer people have fewer resources to buffer against life's ups and downs, but it could also be that people with more stable well-being levels are able to find better-paid jobs. Establishing the causal direction of these relationships would therefore be a valuable area for future research. Nevertheless, our fixed effects filtered regression estimates demonstrate that the results we report are not driven by omitted variables at the individual level such as genetics or personality, and the simple characterisation of these relationships does, in our view, provide a substantial advance in knowledge.

Fourth, people who reported extreme well-being volatility when receiving the app's notification might have chosen to ignore the happiness questionnaire altogether. While this is a valid objection, the consequence of any under-sampling of extreme emotions is that our results are more likely to be underestimated than overestimated. For example, if mothers respond to the Mappiness app's notifications less when experiencing extreme stress during childcare, then we could reasonably expect the extent to which motherhood comes with intense emotional fluctuation to be amplified rather than attenuated in the real world.

Fifth, our results might be sensitive to the length of time used in calculating the within-person variability indicator. Currently, we chose to calculate the within-person variability in happiness on a weekly basis rather than (for example) daily, monthly, or based on specific numbers of responses. As a robustness check, we replaced the intra-week measures of well-being average and volatility with the same quantities calculated across each user's first 14 valid responses, given over 14–28 days, as dependent variables in the OLS regressions. The results, which are reported in Table S4 in *SM*, are qualitatively similar to what had been obtained using the intra-week measures, indicating that the results are not sensitive to how within-person happiness and variability variables are generated.

**Discussion**

Using one of the world's largest ESM data, this paper is one of the first to systematically estimate emotional volatility equations. We discovered, for example, that while larger incomes were significantly associated with greater average happiness, they were strongly predictive of lower emotional volatility over short periods of time. Having children in the household was associated with a more volatile emotional experience, especially for women, who also exhibited more frequent extreme emotional lows, despite the evidence that parents reported similar happiness levels as nonparents on average. Compared to single people, both married and separated individuals experienced a significantly greater emotional volatility.

In addition to the new insights on the valance of different socio-demographic variables in the emotional volatility equation, our study also allowed us to systematically compare the effect size of these factors against each other. Based on Table 2's estimates, we can see, for example, that the estimated effect of having children in the household as a woman on emotional volatility is comparable in size as the effect of being separated. On the other hand, a 1% increase in household income is estimated to only compensate around $\frac{.135}{.021} * 100 \cong 6\%$ of the estimated effect of having children in the household as a woman on emotional volatility. Moreover, we also learned how certain variables such as ill-health are associated not only with low emotional means but also substantially low emotional volatility, thus implying that people with ill-health tend to stay in a stable path of low moods over time. We believe such findings will be of significant interest to people who take emotional volatility into consideration when making daily decisions and life choices. They may also fuel research and debate on whether a life worth living is one that is filled with intense variation in emotions, or one that is stable and content (or, indeed, the extent to which that question has a different answer for each of us).

Our results can also offer new insights into the mechanisms of focusing illusion, i.e., the tendency to exaggerate the impacts of life experiences such as the positive effects of income[43] and parenthood[44] on happiness. Currently, the primary explanation for the empirical evidence of focusing illusion is that people focus too much on conventional beliefs about a particular experience's contribution to happiness when evaluating its effects[45]. However, it could also be that when asked to predict the impact of life experiences that increase emotional volatility, people find one extreme of the wider prospective distribution (either the highs or the lows) more salient than the other. This difference in focus or salience might itself be driven by differential sharing of positive vs negative experiences by others. For example, in the case of parenthood, people may on average focus more on the prospect of moments of great happiness with children and less on the prospect of moments of great distress, even though our asymmetry analysis suggests that extreme negatives are likely to be experienced more frequently. It seems possible that we can improve the accuracy of affective forecasting by asking people explicitly to consider the likely variability and asymmetry in their future experiences — and in the experiences shared by others — and that this could be an interesting subject for future research.

More generally, we believe that, to paint a complete picture of what makes a good life, the science of well-being cannot afford to ignore systematic estimation of the socio-demographic determinants of emotional volatility and asymmetry. In our view, offices of national statistics worldwide should start collecting and analysing this type of emotional well-being data as part of their long-term strategy.

**Methods**

**Sample information.** Participants were 41,023 adults living in the United Kingdom. Of these individuals, 20,343 (49.6%) were women. The mean age was 33.0 years, ranging from 18 to 83. 78.2% worked full-time, 31.5% were married, and 29.0% had children. The median gross annual household income band was GBP £40,000 – £55,999 (mean income cannot be reliably calculated since the top band was unbounded). The user-week mean of daily happiness (scaled 0 – 100) had a mean of 65.6 $\pm$ 16.5 SD. We used the standard deviation of within-person happiness per week as the measure of emotional volatility, which had a mean of 12.8 $\pm$ 9.43 SD across the sample. As our measure of skewness, we used (weekly mean – weekly median) / weekly SD. This quantity had a mean of –0.0780 $\pm$ 0.299 SD.

**Experience sampling procedure.** The Mappiness app was distributed free via Apple's App Store starting in August 2010, and participants were recruited opportunistically. Participants indicated their informed consent to taking part in the study and provided basic demographic information. They were then signalled at random times, within hours and with a frequency they chose (by default, twice between 8am and 10pm). When responding to a signal, users rated their current levels of happiness, relaxation and wakefulness, and were asked where they were, whom they were with, and what they were doing. Subject to user consent, a GPS location was queried from the device and, if outdoors, a photo could be taken (these data are not used in this study). Participants were given simple feedback, summarising their happiness in different contexts, and could opt out of the study at any time.

To generate our measure of within-person variability of happiness per week, we randomly selected one response from each user per day, because we expect intra-day variation to be lower than inter-day variation.

**Experienced well-being and life satisfaction measures.** Our high-frequency experienced wellbeing measure is as follows. During the repeated experience sampling survey, users were asked how far they felt 'Happy', answering on a continuous horizontal sliding scale between 'Not at all' (far left, treated as 0) and 'Extremely' (far right, treated as 100), which began in the central position. Our life satisfaction item was presented once during the app's sign-up process. Users were asked '*How satisfied are you with your life as a whole nowadays?*' and answered by picking a number from 1 – 10 on a horizontal segmented scale.

**Measure of intra-week**

**Measure of the intra-week emotional variability or 'emotional volatility' index, *V*.** We measured the within-person variability of happiness using the following equation:

$$V_{iw} = \sqrt{\frac{\Sigma(h_{iw} - \bar{h}_{iw})^2}{n_w}},$$

where $V_{iw}$ denotes the emotional volatility index for individual *I* at week *w*; $h_{iw}$ is happiness on a 0-100 scale; $\bar{h}_{iw}$ is the average weekly happiness; and $n_w$ is the number of days in the week that the respondent responded to the happiness question. We then standardised $V_{iw}$ to have zero mean and a standard deviation of 1

**Regression analysis.** The linear model of emotional volatility regression equation takes the following form:

$$V_{iw} = \alpha + X'_i \beta + Z'_{iw} \theta + \varepsilon_{iw},$$

where $X'_i$ denotes a vector of socio-demographic variables that do not change over time; and $Z'_{iw}$ is a vector of variables that can vary weekly, including year dummies, previous responses, number of responses this week. Given that people who reported extreme values of happiness are likely to exhibit lower well-being volatility due to the bounded nature of the happiness scale, we included a quadratic term of the happiness average as additional control variables in the emotional volatility regression.

To complement the emotional volatility regression equation, we also estimated the following intra-week emotional average, *M*, regression:

$$M_{iw} = \alpha + X_i'\beta + Z'_{iw}\theta + \varepsilon_{iw},$$

and skewness emotional index, *A*, regression:

$$A_{iw} = \alpha + X_i'\beta + Z'_{iw}\theta + \varepsilon_{iw},$$

where $A_{iw}$ is weekly mean happiness minus the weekly median happiness divided by the weekly standard deviation, i.e., $\frac{Mean_w - Median_w}{V_w}$, and reported the results alongside the well-being volatility estimates. We used ordinary least squares (OLS) with robust standard errors clustered at the individual level to estimate both empirical models.

**Index of within-person variability in daily activities.** The index of within-person variability in daily activities is derived by taking the number of unique activities reported by individual *I* across all their responses in a week and dividing by the total number of responses provided that week. For example, if an individual gave five responses in a week, and reported the same single activity in each of the five, their variability index would be 1 activity / 5 responses = 0.2. If they gave two responses in a week, reporting three activities in the first and two new activities in the second, their index would be 5 activities / 2 responses = 2.5. This index has a mean of 1.23 ± 0.542 SD.

**Fixed effects filtered (FEF) model.**

There are two steps to estimating the FEF (33) model.

Step 1: Compute the person fixed-effects estimator of $\theta$, denoted by $\hat{\theta}$, and the associated residuals, $\hat{\varepsilon}_{iw}$, which is defined by $\hat{\varepsilon}_{iw} = V_{iw} - Z'_{iw}\hat{\theta}$.

Step 2: Compute the within-person averages of these residuals, $\hat{\bar{\varepsilon}}_i = N^{-1}\sum_{w=1}^{N}\hat{\varepsilon}_{iw}$. Regress $\hat{\bar{\varepsilon}}_i$ on $X'_i\beta$ with an intercept to obtain $\hat{\beta}_{FEF}$, where

$$\hat{\beta}_{FEF} = [\sum_{i=1}^{N}(X_i - \bar{X})(X_i - \bar{X})']^{-1}\sum_{i=1}^{N}(X_i - \underline{X})(\hat{\bar{\varepsilon}}_{i_i} - \hat{\bar{\varepsilon}}_i)'.$$

Since the unobserved individual fixed effects is removed from the estimation, $\hat{\beta}_{FEF}$ is free from the usual unobserved heterogeneity bias.

**Data availability statement**

The individual-level data that support the findings of this study are available from the owner and the first author (George MacKerron) but restrictions apply to the availability of these data. This is owing to the terms under which respondents participated, which said: "We won't disclose your data to any third party unless (1) we're required by law to do so, or (2) we do so under a strict contractual agreement with other academic researchers, exclusively for the purpose of academic research at a recognised institution. As a result, the individual-level data are not publicly available. However, the first author is willing to provide the data set on request to academic researchers for replication purposes (and to discuss with them its possible use in other research projects).

**Code availability statement**

All data analysis was conducted using STATA 16. The STATA codes and the publicly-available aggregate data used to generate this study's results can be viewed and downloaded from https://github.com/jawj/happiness-variability.

# References

1. Clark, A. E., & Oswald, A. J. (1994). Unhappiness and unemployment. *Economic Journal*, *104*(424), 648-659.

2. Winkelmann, L., & Winkelmann, R. (1998). Why are the unemployed so unhappy? Evidence from panel data. *Economica*, *65*(257), 1-15.

3. Easterlin, R. A. (1974). Does economic growth improve the human lot? Some empirical evidence. In *Nations and households in economic growth* (pp. 89-125). Academic Press.

4. Easterlin, R. A., McVey, L. A., Switek, M., Sawangfa, O., & Zweig, J. S. (2010). The happiness–income paradox revisited. *Proceedings of the National Academy of Sciences*, *107*(52), 22463-22468.

5. Kahneman, D., & Deaton, A. (2010). High income improves evaluation of life but not emotional well-being. *Proceedings of the National Academy of Sciences*, *107*(38), 16489-16493.

6. Killingsworth, M. A. (2021). Experienced well-being rises with income, even above $75,000 per year. *Proceedings of the National Academy of Sciences*, *118*(4). https://doi.org/10.1073/pnas.2016976118

7. Powdthavee, N., Burkhauser, R. V., & De Neve, J. E. (2017). Top incomes and human well-being: Evidence from the Gallup World Poll. *Journal of Economic Psychology*, *62*, 246-257.

8. Diener, E., Lucas, R., Helliwell, J. F., Schimmack, U., & Helliwell, J. (2009). *Well-being for public policy*. Oxford Positive Psychology.

9. Di Tella, R., MacCulloch, R. J., & Oswald, A. J. (2001). Preferences over inflation and unemployment: Evidence from surveys of happiness. *American Economic Review*, *91*(1), 335-341.

10. Layard, R., Clark, A. E., Cornaglia, F., Powdthavee, N., & Vernoit, J. (2014). What predicts a successful life? A life-course model of well-being. *Economic Journal*, *124*(580), F720-F738.

11. Luechinger, S. (2009). Valuing air quality using the life satisfaction approach. *The Economic Journal*, *119*(536), 482-515.

12. Powdthavee, N., & Van Den Berg, B. (2011). Putting different price tags on the same health condition: Re-evaluating the well-being valuation approach. *Journal of Health Economics*, *30*(5), 1032-1043.

13. Oswald, A. J., & Powdthavee, N. (2008). Death, happiness, and the calculation of compensatory damages. *Journal of Legal Studies*, *37*(S2), S217-S251.

14. Eaton, L. G., & Funder, D. C. (2001). Emotional experience in daily life: valence, variability, and rate of change. *Emotion*, *1*(4), 413.

15. Quoidbach, J., Gruber, J., Mikolajczak, M., Kogan, A., Kotsou, I., & Norton, M. I. (2014). Emodiversity and the emotional ecosystem. *Journal of Experimental Psychology: General*, *143*(6), 2057.

16. Houben, M., Van Den Noortgate, W., & Kuppens, P. (2015). The relation between short-term emotion dynamics and psychological well-being: A meta-analysis. *Psychological Bulletin*, 141(4), 901-930.

17. Dejonckheere, E., Mestdagh, M., Houben, M., Rutten, I., Sels, L., Kuppens, P., & Tuerlinckx, F. (2019). Complex affect dynamics add limited information to the prediction of psychological well-being. *Nature Human Behaviour*, *3*(5), 478-491.

18. Brose, A., Schmiedek, F., Gerstorf, D., & Voelkle, M. C. (2020). The measurement of within-person affect variation. *Emotion*, *20*(4), 677-699.

19. Almeida, D. M., & Kessler, R. C. (1998). Everyday stressors and gender differences in daily distress. Journal of personality and social psychology, 75(3), 670-680.

20. Weigard, A., Loviska, A. M., & Beltz, A. M. (2021). Little evidence for sex or ovarian hormone influences on affective variability. Scientific reports, 11, 1-12.

21. Jachimowicz, J. M., Frey, E. L., Matz, S. C., Jeronimus, B. F., & Galinsky, A. D. (2021). The Sharp Spikes of Poverty: Financial Scarcity Is Related to Higher Levels of Distress Intensity in Daily Life. *Social Psychological and Personality Science*, https://doi.org/10.1177/19485506211060115.

22. Bisconti, T. L., Bergeman, C. S., & Boker, S. M. (2004). Emotional Well-Being in Recently Bereaved Widows: A Dynamical Systems Approach. *The Journals of Gerontology: Series B: Psychological Sciences and Social Sciences*, 59(4), P158-P167

23. Sbarra, D. A., & Emery, R. E. (2005). The emotional sequelae of nonmarital relationship dissolution: Analysis of change and intraindividual variability over time. *Personal Relationships*, 12(2), 213-232.

24. Kerr, M. L., Rasmussen, H. F., Buttitta, K. V., Smiley, P. A., & Borelli, J. L. (2021). Exploring the complexity of mothers' real-time emotions while caregiving. *Emotion*, 21(3), 545-556.

25. Röcke, C., Li, S. C., & Smith, J. (2009). Intraindividual variability in positive and negative affect over 45 days: Do older adults fluctuate less than young adults? *Psychology and aging*, 24(4), 863.

26. Beal, D. J., & Ghandour, L. (2011). Stability, change, and the stability of change in daily workplace affect. *Journal of Organizational Behavior*, *32*(4), 526-546.

27. Lucas, R. E., & Donnellan, M. B. (2007). How stable is happiness? Using the STARTS model to estimate the stability of life satisfaction. *Journal of Research in Personality*, *41*(5), 1091-1098.

28. Kahneman, D., Krueger, A. B., Schkade, D. A., Schwarz, N., & Stone, A. A. (2004). A survey method for characterizing daily life experience: The day reconstruction method. *Science*, *306*(5702), 1776-1780.

29. Csikszentmihalyi, M., & Larson, R. (2014). Validity and reliability of the experience-sampling method. In *Flow and the foundations of positive psychology* (pp. 35-54). Springer, Dordrecht.

30. Bishay, A. (1996). Teacher motivation and job satisfaction: A study employing the experience sampling method. *Journal of Undergraduate Sciences*, *3*(3), 147-155.

31. MacKerron, G & Mourato, S. (2013) Happiness is greater in natural environments. *Global Environmental Change* 23, 992–1000.

32. Seresinhe, C. I., Preis, T., MacKerron, G., & Moat, H. S. (2019). Happiness is greater in more scenic locations. *Scientific Reports*, *9*(1), 1-11.

33. Fujiwara, D., Lawton, R. N., & MacKerron, G. (2017). Experience sampling in and around airports. Momentary subjective wellbeing, airports, and aviation noise in England. *Transportation Research Part D: Transport and Environment*, *56*, 43-54.

34. Bryson, A., & MacKerron, G. (2017). Are you happy while you work? *Economic Journal*, *127*(599), 106-125.

35. Dejonckheere, E., Mestdagh, M., Houben, M., Rutten, I., Sels, L., Kuppens, P., & Tuerlinckx, F. (2019). Complex affect dynamics add limited information to the prediction of psychological well-being. *Nature Human Behaviour*, 3(5), 478-491.

36. Eid, M., & Diener, E. (1999). Intraindividual variability in affect: Reliability, validity, and personality correlates. *Journal of Personality and Social Psychology*, 76(4), 662.

37. Buddelmeyer, H., Hamermesh, D. S., & Wooden, M. (2018). The stress cost of children on moms and dads. *European Economic Review*, *109*, 148-161.

38. Dolan, P., Peasgood, T., & White, M. (2008). Do we really know what makes us happy? A review of the economic literature on the factors associated with subjective well-being. *Journal of Economic Psychology*, *29*(1), 94-122.
39. Cheng, T. C., Powdthavee, N., & Oswald, A. J. (2017). Longitudinal evidence for a midlife nadir in human well-being: Results from four data sets. *Economic Journal*, *127*(599), 126-142.
40. Graham, L., & Oswald, A. J. (2010). Hedonic capital, adaptation and resilience. *Journal of Economic Behavior & Organization*, *76*(2), 372-384.
41. Killingsworth, M. A. (2021). Experienced well-being rises with income, even above $75,000 per year. *Proceedings of the National Academy of Sciences*, *118*(4). https://doi.org/10.1073/pnas.2016976118
42. Pesaran, M. H., & Zhou, Q. (2018). Estimation of time-invariant effects in static panel data models. *Econometric Reviews*, *37*(10), 1137-1171.
43. Kahneman, D., Krueger, A. B., Schkade, D., Schwarz, N., & Stone, A. A. (2006). Would you be happier if you were richer? A focusing illusion. *Science*, *312*(5782), 1908-1910.
44. Powdthavee, N. (2009). Think having children will make you happy? *The Psychologist*, 22(4), 308-310
45. Schkade, D. A., & Kahneman, D. (1998). Does living in California make people happy? A focusing illusion in judgments of life satisfaction. *Psychological Science*, *9*(5), 340-346.

**Acknowledgements:** We are grateful to Andrew Oswald for his invaluable comments on the earlier version of this manuscript and to attendees of the Oxford University's Well-Being Centre seminar series. **Funding**: Neither author received funding to carry out this research. **Authors contributions**: Both authors contributed equally to the project. **Competing interests**: all authors declare no conflicts of interest.

**Supplementary Materials**

Table S1-S4

Fig S1-S3

**Table 1: Overview of selected studies on the socio-demographic determinants of emotional volatility**

| Study | Predictor | Variability measures | Sample size | Method | Duration | Findings |
|---|---|---|---|---|---|---|
| Almeida et al. (1998) | Gender | Daily distress | 166 married couples | Daily diary completed at the end of each day | 42 days | Higher prevalence of daily distress for women compared to men |
| Weigard et al. (2021) | Gender | Affective volatility, emotional inertia, and cyclicity | 142 men (n=30), non-cyclical women (n=28), and oral contraceptive women | Online survey completed at the end of each day | 75 days | Little gender difference in affective variability |
| Jachimowicz et al. (2021) | Financial scarcity | Distress frequency across three different scales | 522 participants | Online survey completed at the end of each day | 6 weeks | Financial scarcity heightens the intensity of daily distress |
| Bisconti et al. (2004) | Widowhood | Daily emotional well-being | 19 older adult widows | Interview | 3 months | Emotional well-being follows an oscillating process that damps across time |
| Sbarra and Emery (2005) | Nonmarital relationship dissolution | Daily emotional well-being, including love, sadness, anger, and relief | 58 young adults | Experienced Sampling Method | 28 days | Linear decline in love and curvilinear patterns for sadness, anger, and relief |
| Kerr et al. (2020) | Motherhood | Positive and negative affects | 136 mothers | Ecological momentary assessment | 10 days | Mothers report higher intensity in positive emotions, and greater variability in both positive and negative emotions when caring for their children. |
| Rocke et al. (2009) | Age | Positive and negative affects | 18 young (20-30 years) and 19 older (70-80 years) adults | Interview | 45 days | Older adults report less positive and negative affects variability than the young |
| Beal and Ghandour (2010) | Employment | Positive and negative affects | 67 employees | Survey completed at the end of each day | 21 days | Positive affect variability was far more predictable by an affective workplace event than negative affect variability |

Table 2. Socio-economic determinants of well-being average, $M$, and volatility, $V$

| Variables | (1) Standardised intra-week emotional average index, $M$ | | (2) Standardised intra-week emotional volatility index, $V$ | |
|---|---|---|---|---|
| | Coeff. | 95% C.I. | Coeff. | 95% C.I. |
| Log of household income | .046 | [.018, .074] | -.021 | [-.036, -.006] |
| **Employment status** | | | | |
| *Base: Employed or self-employed* | | | | |
| Looking after family or home | .051 | [-.074, .176] | -.065 | [-.139, .008] |
| In full-time education | .017 | [-.037, .071] | -.096 | [-.135, -.057] |
| Retired | .012 | [-.385, .410] | -.160 | [-.276, -.043] |
| Long-term sick or disabled | -.214 | [-.365, -.063] | -.110 | [-.212, -.007] |
| Unemployed and seeking work | -.098 | [-.180, -.015] | -.062 | [-.119, -.005] |
| Other | -.042 | [-.150, .065] | -.076 | [-.151, -.0003] |
| **Gender × children in household** | | | | |
| *Base: Male, no children* | | | | |
| Female, no children | -.031 | [-.070, .007] | .094 | [.069, .118] |
| Female, with children | .008 | [-.050, .066] | .135 | [.099, .170] |
| Male, with children | .029 | [-.026, .084] | .040 | [.003, .077] |
| **Marital status** | | | | |
| *Base: Single* | | | | |
| Divorced | .017 | [-.081, .114] | .089 | [.032, .145] |
| Married | .059 | [.014, .103] | .052 | [.021, .083] |
| Separated | -.095 | [-.191, .002] | .107 | [.037, .178] |
| Widowed | .052 | [-.237, .340] | .042 | [-.092, .176] |
| **Number of adults in the household** | | | | |
| 2 | .044 | [-.004, .092] | .067 | [.038, .097] |
| 3 | .009 | [-.052, .069] | .043 | [.002, .084] |
| 4+ | -.003 | [-.067, .062] | .053 | [.014, .091] |
| **Age (at first response of week)** | | | | |
| Age | -.024 | [-.036, -.012] | -.025 | [-.033, -.018] |
| Age squared | .0003 | [.0002, .0005] | .0001 | [.0001, .0003] |
| **Self-reported health** | | | | |
| *Base: Excellent* | | | | |
| Poor | -.701 | [-.825, -.577] | -.004 | [-.089, .082] |
| Fair | -.546 | [-.623, -.469] | -.042 | [-.089, .0056] |
| Good | -.324 | [-.377, -.270] | -.074 | [-.106, -.042] |
| Very good | -.114 | [-.165, -.063] | -.079 | [-.109, -.049] |
| **Home region** | | | | |
| *Base: London* | | | | |
| North East | .148 | [.043, .253] | .137 | [.061, .213] |
| North West | .081 | [.009, .152] | .072 | [.025, .119] |
| Yorkshire and The Humber | .074 | [-.002, .150] | .060 | [.013, .107] |
| East Midlands | .159 | [.054, .264] | .075 | [.012, .137] |
| West Midlands | .013 | [-.096, .122] | .044 | [-.015, .104] |
| East of England | .027 | [-.052, .105] | .054 | [.002, .105] |
| South East | .052 | [.002, .103] | .036 | [-.002, .074] |
| South West | -.001 | [-.083, .081] | .006 | [-.041, .054] |
| Northern Ireland | -.075 | [-.247, .096] | -.006 | [-.092, .081] |
| Scotland | .082 | [-.012, .177] | .048 | [-.018, .114] |
| Wales | .053 | [-.031, .137] | .055 | [-.012, .121] |
| Unknown location | -.046 | [-.0913, .0003] | .030 | [-.001, .062] |
| **Year (at first response of week)** | | | | |
| *Base: 2010* | | | | |
| 2011 | .028 | [.004, .051] | -.111 | [-.129, -.093] |
| 2012 | -.039 | [-.076, -.002] | -.124 | [-.151, -.098] |
| 2013 | -.097 | [-.135, -.059] | -.116 | [-.144, -.088] |
| 2014 | -.119 | [-.175, -.063] | -.127 | [-.165, -.088] |
| 2015 | -.118 | [-.199, -.037] | -.125 | [-.177, -.073] |
| 2016 | -.113 | [-.196, -.031] | -.123 | [-.180, -.066] |
| **Prior responses** | | | | |
| Count | .0002 | [.0000533, .000268] | -.0003 | [-.000413, -.000267] |
| Count squared | -3.38e-08 | [-7.32e-08, 5.60e-09] | 4.85e-08 | [2.18e-08, 7.52e-08] |
| **Number of responses in the $w^{th}$ week** | | | | |
| *Base: 2* | | | | |
| 3 | .028 | [.013, .043] | .114 | [.099, .129] |
| 4 | .039 | [.022, .056] | .156 | [.141, .171] |
| 5 | .061 | [.041, .081] | .178 | [.162, .194] |
| 6 | .086 | [.064, .107] | .185 | [.168, .202] |
| 7 | .130 | [.103, .157] | .180 | [.160, .200] |
| **Intra-week emotional average, *M*** | | | | |
| Standardised mean | | | -.326 | [-.335, -.318] |
| Standardised mean squared | | | -.093 | [-.099, -.088] |
| Constant | .012 | [-.310, .334] | .864 | [.666, 1.06] |
| N (user-weeks) | 338,447 | | 338,447 | |
| Users | 41,023 | | 41,023 | |
| R-squared | 0.051 | | 0.155 | |
| Adjusted R-squared | 0.051 | | 0.155 | |
| F-statistic | 20.17 | | 226.835 | |

**Note**: We used ordinary least squares (OLS) regressions to estimate these models. Standard errors are clustered at the individual personal identification level. The sample includes only participants with at least two responses per week. If there are multiple responses from a user in a single day, one response is selected at random to generate the mean and variability variables. The dependent variables are standardised to have a mean of 0 and a standard deviation of 1.

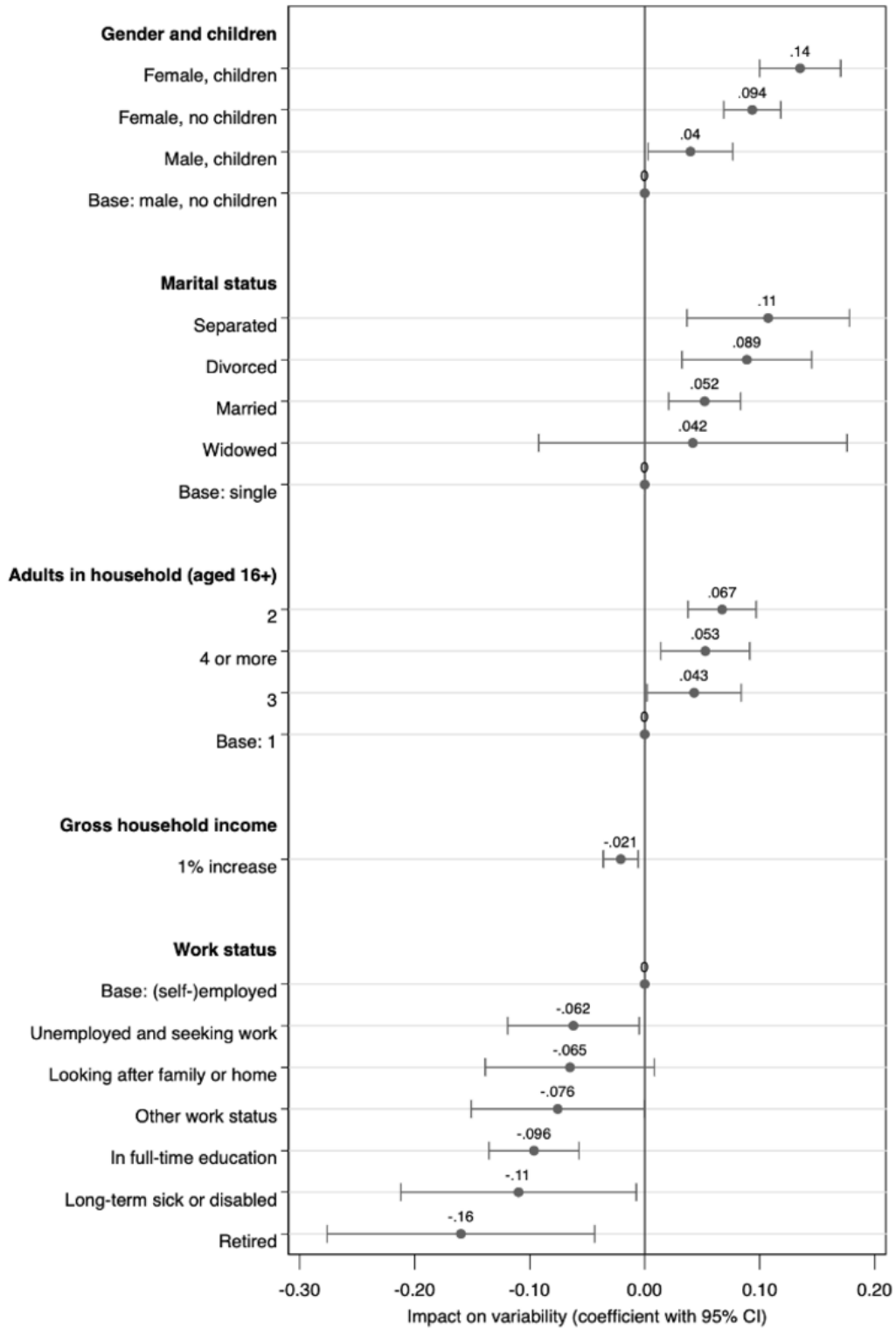

**Fig. 1.** This figure plots socio-economic variables according to their *z* scores for intra-week emotional volatility index, *V*. Each *z* score is based on the estimates obtained from running the emotional volatility regression equation reported in Column 2 of Table 1. 95% confidence intervals are displayed here.

**Table 3: Socio-economic determinants of well-being skewness or asymmetry, *A***

|  | Standardised intra-week emotional skewness, $A$ | |
|---|---|---|
| **Variables** | Coeff. | 95% C.I. |
| Log of household income | .002 | [-.007, .010] |
| **Employment status** | | |
| *Base: Employed or self-employed* | | |
| Looking after family or home | -.017 | [-.063, .029] |
| In full-time education | .022 | [.002, .042] |
| Retired | .019 | [-.055, .095] |
| Long-term sick or disabled | -.023 | [-.074, .027] |
| Unemployed and seeking work | -.006 | [-.036, .024] |
| Other | .006 | [-.033, .045] |
| **Gender × children in household** | | |
| *Base: Male, no children* | | |
| Female, no children | -.010 | [-.023, .004] |
| Female, with children | -.049 | [-.069, -.029] |
| Male, with children | -.017 | [-.037, .004] |
| **Marital status** | | |
| *Base: Single* | | |
| Divorced | -.014 | [-.045, .018] |
| Married | -.004 | [-.022, .013] |
| Separated | -.017 | [-.050, .016] |
| Widowed | .011 | [-.102, .124] |
| **Number of adults in the household** | | |
| 2 | -.015 | [-.032, .002] |
| 3 | -.002 | [-.025, .021] |
| 4+ | -.015 | [-.037, .007] |
| **Age (at first response of week)** | | |
| Age | .0004 | [-.004, .005] |
| Age squared | 7.40e-06 | [-.00004, .00006] |
| **Self-reported health** | | |
| *Base: Excellent* | | |
| Poor | -.049 | [-.097, -.003] |
| Fair | -.017 | [-.042, .008] |
| Good | -.014 | [-.033, .005] |
| Very good | .005 | [-.013, .022] |
| **Home region** | | |
| *Base: London* | | |
| North East | -.007 | [-.0495, .0364] |
| North West | -.021 | [-.0464, .00492] |
| Yorkshire and The Humber | .002 | [-.0253, .0299] |
| East Midlands | -.022 | [-.0586, .0155] |
| West Midlands | -.009 | [-.0389, .0218] |
| East of England | -.013 | [-.0432, .0166] |
| South East | .002 | [-.0186, .0224] |
| South West | -.011 | [-.0388, .017] |
| Northern Ireland | -.005 | [-.0554, .0448] |
| Scotland | .004 | [-.0288, .0375] |
| Wales | -.029 | [-.0776, .0198] |
| Unknown location | -.003 | [-.0196, .0144] |
| **Year (at first response of week)** | | |
| *Base: 2010* | | |
| 2011 | .023 | [.0118, .0346] |
| 2012 | .004 | [-.0126, .0215] |
| 2013 | -.018 | [-.0353, .00029] |
| 2014 | .005 | [-.018, .0281] |
| 2015 | .005 | [-.0277, .0384] |
| 2016 | -.009 | [-.0431, .026] |
| **Prior responses** | | |
| Count | .0001 | [.0001, .0002] |
| Count squared | -2.27e-08 | [-3.50e-08, -1.05e-08] |
| **Number of responses in the $w^{th}$ week** | | |
| *Base: 2* | | |
| 3 | .102 | [.0866, .116] |
| 4 | .100 | [.0882, .112] |
| 5 | .034 | [.0211, .0472] |
| 6 | .035 | [.024, .0453] |
| 7 | .000 | [.000, 0.000] |
| **Intra-week emotional average, *M*** | | |
| Standardised mean | -.126 | [-.134, -.118] |
| Standardised mean squared | .013 | [.008, .019] |
| Constant | -.098 | [-.209, .012] |
| N (user-weeks) | 285,947 | |
| Users | 36,893 | |
| R-squared | 0.019 | |
| Adjusted R-squared | 0.019 | |
| F-statistic | 45.83 | |

**Note**: We used ordinary least squares (OLS) regressions to estimate these models. Standard errors are clustered at the individual personal identification level. Weekly asymmetry in happiness is calculated using the following equation: $\frac{Mean_w - Median_w}{Standard\ Deviation_w}$, which captures the degree of 'skewness' in the experienced happiness per distribution. The sample is reduced to include only the sample with at least three observations per week. Other controls include dummy variables representing home regions, year, number of responses in the previous week and its square, number of responses in the corresponding week. The dependent variable is standardised to have a mean of 0 and a standard deviation of 1.

**Descriptive Statistics, Regression Results, and Robustness Checks**

**Figs. S1A-C: Distributions of emotional average, volatility, and skewness**

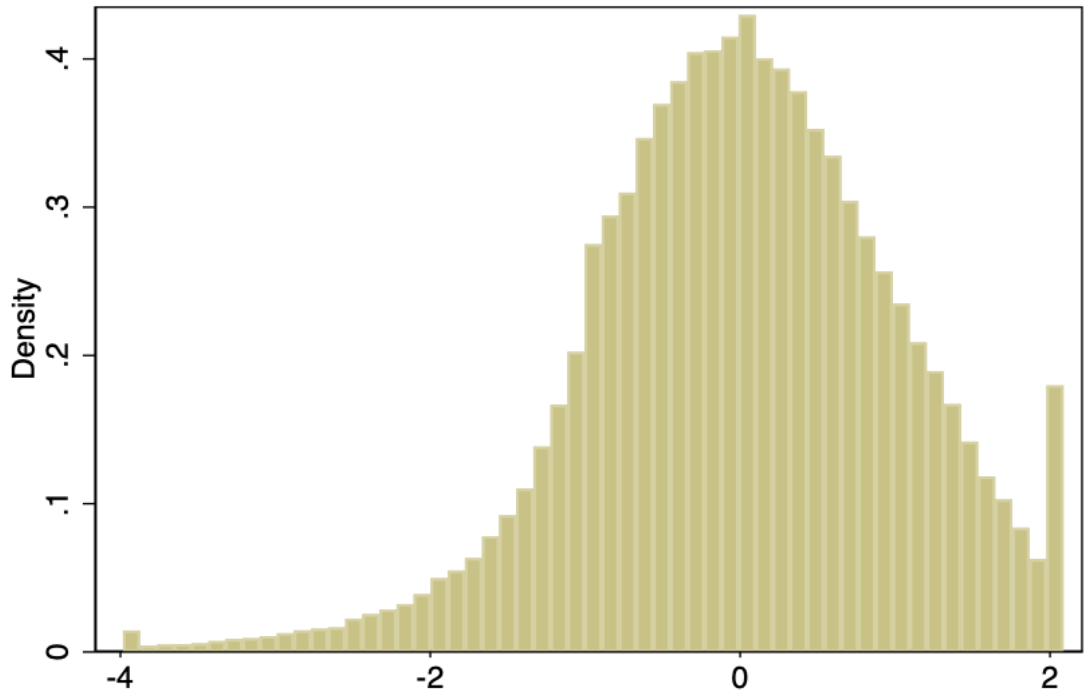

Fig. S1A: Distribution of standardised intra-week emotional average index, *M*

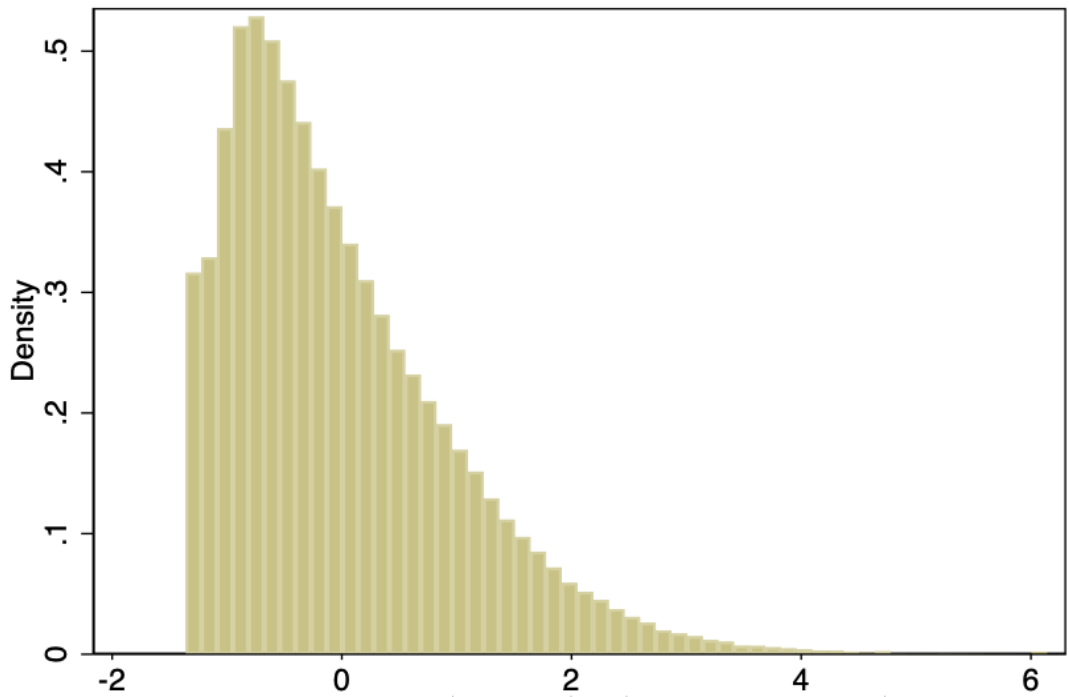

Fig S1B: Distribution of standardised intra-week emotional volatility index, *V*

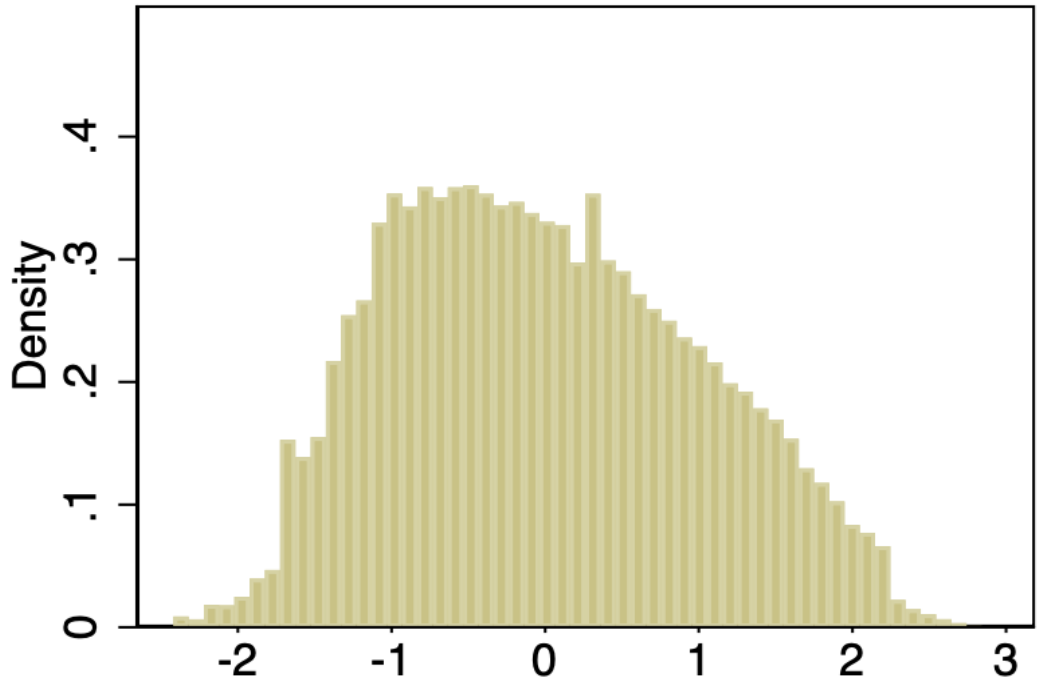

**Fig S1C**: Distribution of standardised intra-week emotional skewness index, *A*

**Fig. S2: Plot of socio-economic variables according to their *z* scores for intra-week emotional average, *M*, and volatility, *V***

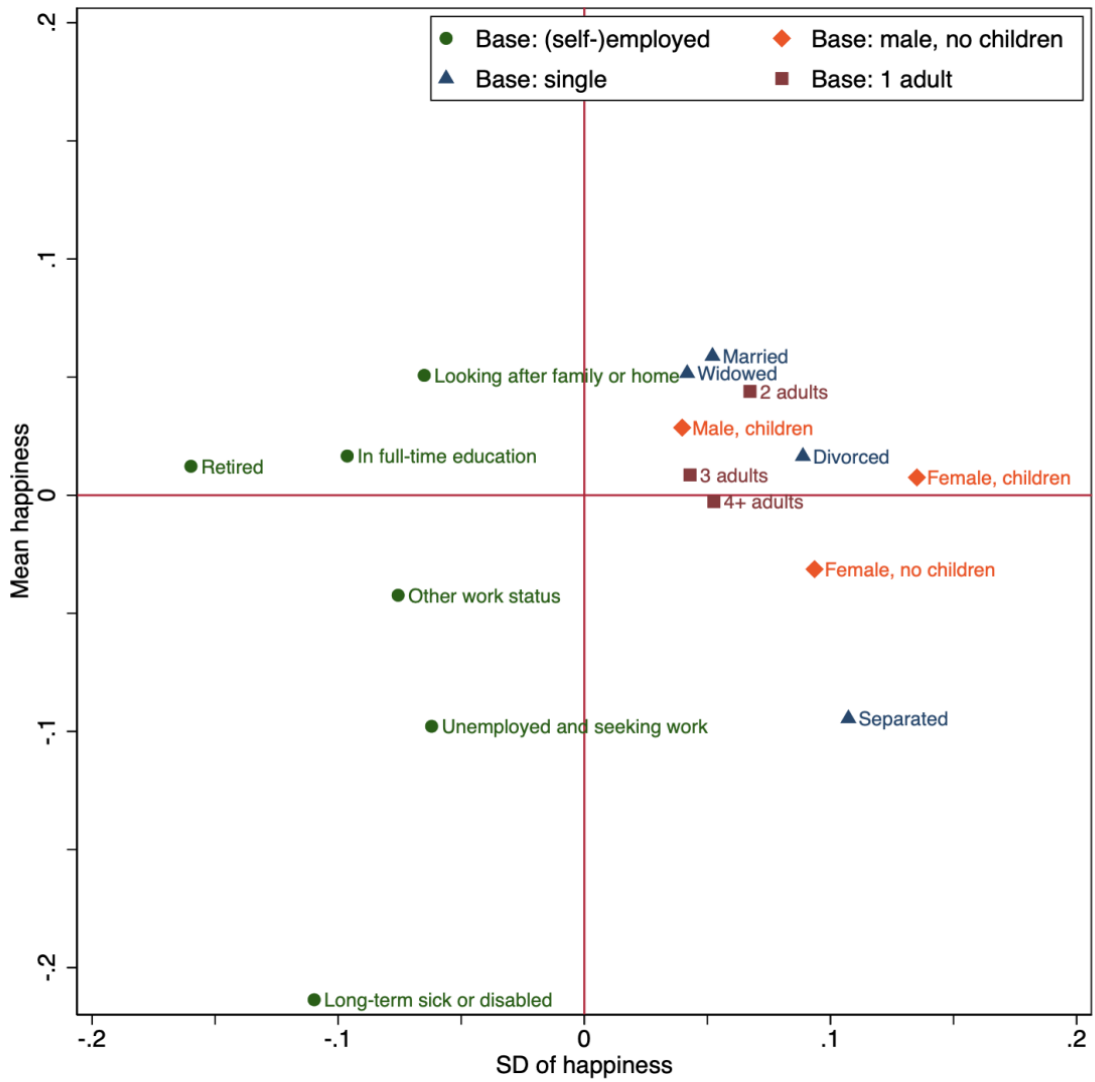

**Note:** $M$ = Mean of happiness; $V$ = SD of happiness.

**Figs. S3A-B: An illustration of two hypothetical individuals with the same emotional volatility but different emotional skewness index**

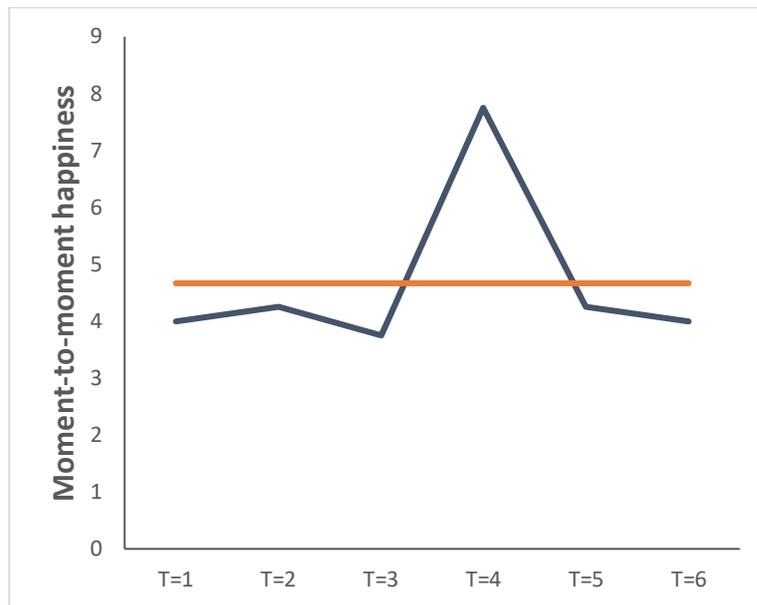

**Fig. S4A:** Positively skewed

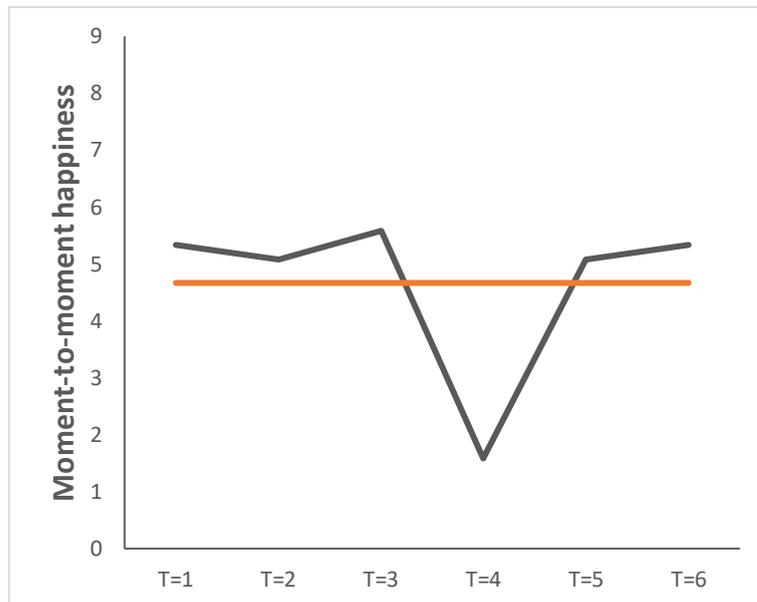

**Fig. S4B:** Negatively skewed

**Note:** The data plotted in Figs. S3A and S3B share the same mean (4.67) and standard deviation (1.52). However, the emotional skewness is different across the two distributions. The skewness indices are 0.36 in Fig. S3A and -0.36 in Fig. S3B.

**Table S1: Standardised intra-week emotional volatility index, *V*, with within-person variability in activities as a control variable**

|  | Model 1 | | Model 2 | |
| --- | --- | --- | --- | --- |
| **Variables** | **Coeff.** | **95% C.I.** | **Coeff.** | **95% C.I.** |
| Log of household income | -.021 | [-.036, -.006] | -.020 | [-.035, -.005] |
| **Employment status** | | | | |
| *Base: Employed or self-employed* | | | | |
| Looking after family or home | -.065 | [-.139, .008] | -.071 | [-.145, .004] |
| In full-time education | -.096 | [-.135, -.057] | -.099 | [-.138, -.059] |
| Retired | -.160 | [-.276, -.044] | -.160 | [-.273, -.047] |
| Long-term sick or disabled | -.110 | [-.212, -.007] | -.117 | [-.218, -.015] |
| Unemployed and seeking work | -.062 | [-.119, -.005] | -.066 | [-.124, -.009] |
| Other | -.076 | [-.151, -.000] | -.085 | [-.160, -.010] |
| **Gender × children in household** | | | | |
| *Base: Male, no children* | | | | |
| Female, no children | .093 | [.069, .118] | .086 | [.061, .111] |
| Female, with children | .135 | [.099, .170] | .128 | [.093, .164] |
| Male, with children | .039 | [.003, .077] | .039 | [.002, .075] |
| **Marital status** | | | | |
| *Base: Single* | | | | |
| Divorced | .089 | [.033, .145] | .089 | [.034, .146] |
| Married | .052 | [.021, .083] | .053 | [.022, .084] |
| Separated | .107 | [.037, .178] | .104 | [.033, .175] |
| Widowed | .042 | [-.092, .176] | .037 | [-.095, .168] |
| **Number of adults in the household** | | | | |
| *Base: 1* | | | | |
| 2 | .067 | [.038, .097] | .068 | [.038, .097] |
| 3 | .043 | [.002, .084] | .043 | [.002, .084] |
| 4+ | .053 | [.014, .091] | .052 | [.013, .091] |
| **Age (at first response of week)** | | | | |
| Age | -.025 | [-.032, -.018] | -.024 | [-.032, -.0169] |
| Age squared | .0002 | [.0001, .0003] | .0002 | [.0001, .0003] |
| **Self-reported health** | | | | |
| *Base: Excellent* | | | | |
| Poor | -.004 | [-.089, .082] | -.014 | [-.099, .073] |
| Fair | -.042 | [-.089, .0056] | -.046 | [-.093, .001] |
| Good | -.074 | [-.106, -.042] | -.076 | [-.108, -.044] |
| Very good | -.079 | [-.109, -.049] | -.081 | [-.112, -.051] |
| **Intra-week emotional average, *M*** | | | | |
| Standardised mean | -.326 | [-.335, -.318] | -.332 | [-.341, -.324] |
| Standardised mean squared | -.093 | [-.099, -.088] | -.093 | [-.098, -.087] |
| Within-person activity variability |  |  | .155 | [.115, .194] |
| Within-person activity variability squared |  |  | -.021 | [-.032, -.010] |
| Constant | .864 | [.666, 1.060] | .676 | [.474, .877] |

| | | |
|---|---|---|
| N (user-weeks) | 338447 | 338447 |
| Users | 41023 | 41023 |
| R-squared | 0.155 | 0.157 |
| Adjusted R-squared | 0.155 | 0.157 |
| F-statistic | 226.835 | 221.779 |

**Note:** Other control variables include home region dummies, the number of responses in the $w^{th}$ week, prior responses and its squared.

## Table S2: Life satisfaction regression

| Variables | Coeff. | 95% C.I. |
|---|---|---|
| Log of household income | .166 | [.145, .187] |
| **Employment status** | | |
| *Base: Employed or self-employed* | | |
| Looking after family or home | -.038 | [-.153, .077] |
| In full-time education | .200 | [.148, .252] |
| Retired | .339 | [.140, .539] |
| Long-term sick or disabled | -.547 | [-.755, -.339] |
| Unemployed and seeking work | -.937 | [-1.040, -.837] |
| Other | -.066 | [-.190, .057] |
| **Gender × children in household** | | |
| *Base: Male, no children* | | |
| Female, no children | .046 | [.012, .079] |
| Female, with children | .032 | [-.019, .083] |
| Male, with children | .026 | [-.022, .073] |
| **Marital status** | | |
| *Base: Single* | | |
| Divorced | .053 | [-.021, .128] |
| Married | .286 | [.243, .329] |
| Separated | -.266 | [-.368, -.164] |
| Widowed | .169 | [-.095, .434] |
| **Number of adults in the household** | | |
| *Base: 1* | | |
| 2 | .221 | [.181, .261] |
| 3 | -.018 | [-.071, .036] |
| 4 or more | .034 | [-.025, .093] |
| **Age (at first response of week)** | | |
| Age | -.058 | [-.068, -.047] |
| Age squared | .001 | [.0005, .0008] |
| **Self-reported health** | | |
| *Base: Excellent* | | |
| Poor | -2.390 | [-2.550, -2.220] |
| Fair | -1.690 | [-1.760, -1.620] |
| Good | -.987 | [-1.030, -.942] |
| Very good | -.384 | [-.426, -.342] |
| **Home region** | | |
| *Base: London* | | |
| North East | .044 | [-.069, .157] |
| North West | -.041 | [-.109, .027] |
| Yorkshire and The Humber | .032 | [-.044, .108] |
| East Midlands | .025 | [-.059, .110] |
| West Midlands | .042 | [-.035, .119] |
| East of England | -.028 | [-.097, .041] |
| South East | .026 | [-.028, .080] |
| South West | -.014 | [-.086, .058] |
| Northern Ireland | -.156 | [-.311, .0002] |
| Scotland | .034 | [-.042, .110] |
| Wales | .003 | [-.093, .098] |
| Unknown/non-UK | .052 | [.010, .094] |
| Constant | 6.320 | [6.030, 6.600] |
| N (user-weeks) | 338447 | |
| Users | 41023 | |
| R-squared | .043 | |
| Adjusted R-squared | .043 | |
| F-statistic | 14.903 | |

**Note:** Life satisfaction is standardised to have a mean of 0 and a standard deviation of 1.

**Table S3: Fixed effects filtered model**

| Variables | (1) Standardised Intra-week emotional average index, $M$ | | (2) Standardised Intra-week emotional volatility index, $V$ | |
|---|---|---|---|---|
| | **Coeff.** | **95% C.I.** | **Coeff.** | **95% C.I.** |
| Log of household income | .050 | [.038, .061] | -.017 | [-.028, -.005] |
| **Employment status** | | | | |
| *Base: Employed or self-employed* | | | | |
| Looking after family or home | .030 | [-.034, .094] | -.024 | [-.086, .037] |
| In full-time education | .026 | [-.007, .058] | -.138 | [-.170, -.107] |
| Retired | .087 | [-.039, .212] | .071 | [-.040, .182] |
| Long-term sick or disabled | -.181 | [-.295, -.066] | -.028 | [-.119, .064] |
| Unemployed and seeking work | -.111 | [-.164, -.058] | -.068 | [-.117, -.019] |
| Other | -.020 | [-.083, .043] | -.076 | [-.142, -.011] |
| **Gender × children in household** | | | | |
| *Base: Male, no children* | | | | |
| Female, no children | -.028 | [-.049, -.007] | .063 | [.043, .083] |
| Female, with children | -.013 | [-.043, .017] | .146 | [.116, .175] |
| Male, with children | .019 | [-.008, .046] | .058 | [.031, .085] |
| **Marital status** | | | | |
| *Base: Single* | | | | |
| Divorced | .056 | [.013, .099] | .162 | [.120, .203] |
| Married | .082 | [.058, .107] | .067 | [.042, .092] |
| Separated | -.066 | [-.125, -.006] | .150 | [.098, .201] |
| Widowed | .025 | [-.135, .185] | .137 | [.010, .263] |
| **Number of adults in the household** | | | | |
| *Base: 1* | | | | |
| 2 | .051 | [.028, .107] | .072 | [.050, .094] |
| 3 | -.008 | [-.038, .023] | .042 | [.012, .072] |
| 4+ | -.001 | [-.033, .032] | .045 | [.013, .078] |
| **Age (at first response of week)** | | | | |
| Age | -.026 | [-.033, -.020] | -.038 | [-.044, -.032] |
| Age squared | .000 | [.0002, .0004] | .000 | [.0002, .0004] |
| **Self-reported health** | | | | |
| *Base: Excellent* | | | | |
| Poor | -.774 | [-.859, -.688] | -.058 | [-.136, .019] |
| Fair | -.538 | [-.575, -.502] | -.143 | [-.179, -.107] |
| Good | -.349 | [-.374, -.323] | -.150 | [-.175, -.124] |
| Very good | -.135 | [-.159, -.110] | -.126 | [-.150, -.102] |
| **Home region** | | | | |
| *Base: London* | | | | |
| North East | .166 | [.104, .229] | .218 | [.160, .276] |
| North West | .086 | [.048, .124] | .098 | [.062, .135] |
| Yorkshire and The Humber | .098 | [.056, .141] | .119 | [.076, .162] |
| East Midlands | .102 | [.055, .148] | .115 | [.070, .161] |
| West Midlands | .129 | [.085, .173] | .156 | [.113, .199] |
| East of England | .059 | [.020, .098] | .091 | [.052, .130] |
| South East | .070 | [.040, .101] | .086 | [.056, .115] |
| South West | .066 | [.027, .106] | .075 | [.037, .113] |
| Northern Ireland | -.022 | [-.103, .059] | .085 | [-.006, .175] |
| Scotland | .125 | [.082, .168] | .111 | [.068, .154] |
| Wales | .123 | [.070, .175] | .120 | [.065, .174] |
| Unknown location | -.047 | [-.073, -.020] | .052 | [.027, .078] |
| **Year (at first response of week)** | | | | |
| *Base: 2010* | | | | |
| 2011 | .000 | [-.024, .046] | -.052 | [-.073, -.031] |
| 2012 | -.040 | [-.087, .041] | -.028 | [-.069, .012] |
| 2013 | -.109 | [-.192, -.014] | -.003 | [-.060, .055] |
| 2014 | -.090 | [-.235, .002] | .019 | [-.052, .090] |
| 2015 | -.107 | [-.263, .063] | .067 | [-.022, .156] |
| 2016 | -.091 | [-.278, .132] | .115 | [.003, .226] |
| **Prior responses** | | | | |
| Count | .0001 | [-.0004, .0005] | -.0002 | [-.0003, -.0002] |
| Count squared | -1.92e-08 | [-3.94e-08, -3.50e-09] | 3.44e-08 | [2.72e-08, 4.16e-08] |
| **Number of responses in the $w^{th}$ week** | | | | |
| *Base: 2* | | | | |
| 3 | -.001 | [-.013, .010] | .111 | [.097, .124] |
| 4 | -.010 | [-.021, .002] | .164 | [.151, .177] |
| 5 | -.009 | [-..021, .003] | .196 | [.182, .209] |
| 6 | -.004 | [-.017, .008] | .214 | [.200, .227] |
| 7 | .011 | [-.003, .025] | .227 | [.213, .242] |
| **Intra-week emotional average, $M$** | | | | |
| Standardised mean | | | -.611 | [-.623, -.599] |
| Standardised mean squared | | | -.210 | [-.216, -.204] |
| N (user-weeks) | 338447 | | 338447 | |
| Users | 41023 | | 41023 | |

**Table S4: Recalculation of the within-person emotional mean and volatility**

| Variables | Standardised emotional average index from the first 28 days, M | | Standardised emotional volatility index from the first 28 days, V | |
|---|---|---|---|---|
| | Coeff. | 95% C.I. | Coeff. | 95% C.I. |
| Log of household income | .028 | [.009, .047] | -.045 | [-.063, -.026] |
| **Employment status** | | | | |
| *Base: Employed or self-employed* | | | | |
| Looking after family or home | .066 | [-.036, .168] | -.015 | [-.107, .078] |
| In full-time education | -.025 | [-.073, .024] | -.182 | [-.230, -.134] |
| Retired | .265 | [.0603, .469] | -.136 | [-.283, .012] |
| Long-term sick or disabled | -.089 | [-.266, .088] | .008 | [-.139, .156] |
| Unemployed and seeking work | -.115 | [-.200, -.031] | -.089 | [-.162, -.017] |
| Other | .016 | [-.084, .115] | -.032 | [-.137, .072] |
| **Gender × children in household** | | | | |
| *Base: Male, no children* | | | | |
| Female, no children | -.038 | [-.068, -.008] | .133 | [.104, .163] |
| Female, with children | .039 | [-.006, .085] | .222 | [.178, .266] |
| Male, with children | .032 | [-.010, .075] | .0314 | [-.011, .074] |
| **Marital status** | | | | |
| *Base: Single* | | | | |
| Divorced | .085 | [.018, .151] | .169 | [.107, .232] |
| Married | .116 | [.078, .155] | .033 | [-.004, .071] |
| Separated | .029 | [-.058, .117] | .230 | [.150, .310] |
| Widowed | .038 | [-.233, .310] | .154 | [-.058, .367] |
| **Number of adults in the household** | | | | |
| 2 | .077 | [.041, .112] | .093 | [.059, .127] |
| 3 | .0002 | [-.048, .048] | .090 | [.044, .136] |
| 4+ | .061 | [.007, .114] | .047 | [-.004, .098] |
| **Age (at first response of week)** | | | | |
| Age | -.041 | [-.051, -.030] | -.031 | [-.040, -.022] |
| Age squared | .0005 | [.0004, .0006] | .0002 | [.0001, .0003] |
| **Self-reported health** | | | | |
| *Base: Excellent* | | | | |
| Poor | -.924 | [-1.060, -.783] | .176 | [.057, .296] |
| Fair | -.672 | [-.730, -.614] | .037 | [-.020, .094] |
| Good | -.413 | [-.454, -.373] | -.056 | [-.096, -.016] |
| Very good | -.175 | [-.213, -.136] | -.089 | [-.126, -.052] |
| **Home region** | | | | |
| *Base: London* | | | | |
| North East | .207 | [.107, .306] | .200 | [.113, .287] |
| North West | .128 | [.069, .187] | .112 | [.056, .169] |
| Yorkshire and The Humber | .136 | [.072, .200] | .125 | [.058, .192] |
| East Midlands | .152 | [.084, .222] | .122 | [.053, .191] |
| West Midlands | .145 | [.077, .213] | .126 | [.058, .194] |
| East of England | .123 | [.063, .183] | .055 | [-.004, .114] |
| South East | .110 | [.064, .156] | .069 | [.025, .115] |
| South West | .094 | [.033, .154] | .071 | [.011, .130] |
| Northern Ireland | .029 | [-.104, .163] | .040 | [-.093, .174] |
| Scotland | .136 | [.070, .202] | .046 | [-.019, .112] |
| Wales | .107 | [.028, .187] | .130 | [.045, .216] |
| Unknown location | -.041 | [-.079, -.002] | .00686 | [-.030, .0437] |
| **Weekly emotional average, *M*** | | | | |
| Standardised mean | | | -.236 | [-.249, -.224] |
| Standardised mean squared | | | -.072 | [-.079, -.064] |
| Constant | .664 | [.397, .931] | 1.16 | [.903, 1.410] |
| N | 22365 | | 22365 | |
| R-squared | .065 | | .104 | |
| Adjusted R-squared | .063 | | .102 | |
| F-statistic | 28.79 | | 56.02 | |

**Note:** Dependent variables are calculated using the data taken from the first 14 responses (limiting to 1 per day) within 28 days. An additional control includes dummy variables representing days to reach 14 responses.